\input{aipcheck}

\documentclass[
    ,final            % use final for the camera ready runs
  ]
  {aipproc}

\usepackage{amsmath}
\usepackage{amssymb}
\usepackage{epsfig}

\usepackage{fontenc}
\usepackage{times}
\usepackage{mathptmx}
\usepackage{graphicx}
\usepackage{subfigure}

\newcommand{\be}{\begin{equation}}
\newcommand{\ee}{\end{equation}}

\newcommand{\startitem}{\begin{itemize}}
\newcommand{\stopitem}{\end{itemize}}
\newcommand{\bs}{\begin{split}}
\newcommand{\es}{\end{split}}

\newcommand{\KET}[1]{\ensuremath{|#1\rangle}}

\newcommand{\VEC}[1]{\ensuremath{\mbox{\boldmath{$#1$}}}}

\newcommand{\bea}{\begin{eqnarray}}
\newcommand{\eea}{\end{eqnarray}}

\layoutstyle{6x9}

\begin{document}

\title{$p\bar{p} \rightarrow \Lambda_c \bar{\Lambda}_c$ within a Handbag Picture -- \\
  Cross Section and Spin Observables
\footnote{Talk given by A.T.~Goritschnig at the {\em 18th
International Spin Physics Symposium (SPIN 2008)},\hfill\break
Charlottesville, USA, Oct. 2008.}
}


\author{A.T.~Goritschnig}{
  address={Institut f\"ur Physik, Karl-Franzens-Universit\"at Graz, Austria}
}

\author{P.~Kroll}{
  address={Fachbereich Physik, Universit\"at Wuppertal, Germany}
}

\author{W.~Schweiger}{
  address={Institut f\"ur Physik, Karl-Franzens-Universit\"at Graz, Austria} % additional visiting address
}

\begin{abstract}
We study the process $p\bar{p} \rightarrow \Lambda_c \bar{\Lambda}_c$
within the generalized parton picture.
Our starting point is the double handbag diagram
which factorizes into soft generalized parton distributions
and a hard subprocess amplitude for $u \bar{u} \rightarrow c \bar{c}$.
Our cross-section predictions may become interesting in view of
forthcoming experiments at FAIR in Darmstadt.
\end{abstract}

\maketitle

%%%%%%%%%%%%%%%%%%%%%%%%%%%%%%%%%%%%%%%%%%%%
%% MAINMATTER
%%%%%%%%%%%%%%%%%%%%%%%%%%%%%%%%%%%%%%%%%%%%

\section{Introduction}
For exclusive hadronic reactions which require the production of
heavy quark-antiquark pairs QCD perturbation theory has a good
chance to provide a substantial part of the process amplitude
already at moderately large energies.
The reasons are that the mass of the heavy quark can serve as a hard scale
and that certain reaction mechanisms can immediately be ruled out,
since the heavy-flavor content of the quark sea in the proton is small.
\begin{center}
\begin{figure}[h]
%\begin{center}
\epsfig{file=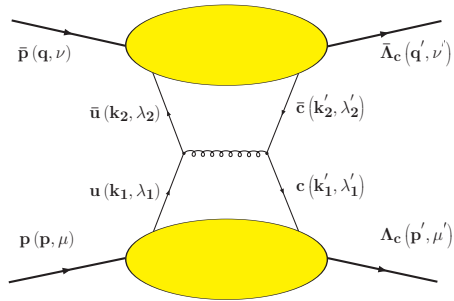,width=6cm,clip=}
\caption{Double-handbag contribution to $p\bar{p} \rightarrow \Lambda_c\bar{\Lambda}_c$}
\label{Fig:Factorization}
%\end{center}
\end{figure}
\end{center}

For the process $p \bar{p} \rightarrow \Lambda_c \bar{\Lambda}_c$
the reaction mechanism which we assume to be dominant in the
forward hemisphere for energies well above production threshold is
depicted in Fig.~\ref{Fig:Factorization}. We analyze it in terms
of generalized parton distributions~\cite{Radyushkin:1997ki},
\cite{Ji:1996nm}.

If the dynamical mechanism underlying $p  \bar{p} \, \rightarrow
\, \Lambda_c \bar{\Lambda}_c$ scattering is given by
Fig.~\ref{Fig:Factorization}, it can be
shown~\cite{Goritschnig:2008} -- in analogy to Compton
scattering~\cite{Diehl:1998kh}, \cite{Diehl:1999tr} -- that the
corresponding scattering amplitude factorizes into soft hadronic
matrix elements for the $p \rightarrow \Lambda_c$ (and $\bar{p}
\rightarrow \bar{\Lambda}_c$) transition and a hard subprocess
amplitude for $u \bar{u} \rightarrow c \bar{c}$. The subprocess
amplitude can be calculated perturbatively, whereas the hadronic
matrix elements describe the non-perturbative transition of the
(anti-)proton to the (anti-)$\Lambda_c$ by emission of a
($\bar{u}$) $u$ quark and reabsorption of a ($\bar{c}$) $c$ quark.
It is just the hadronic matrix elements which give rise to
generalized parton distribution functions (GPDs) and to form
factors (FFs), which are essentially $1/x$ moments of the
GPDs~\cite{Radyushkin:1997ki}, \cite{Ji:1996nm} after decomposing
them into their covariant structures. At this point eight FFs have
to be introduced. If it is assumed that contributions from
non-zero orbital angular momenta play a minor role in the proton
and $\Lambda_c$ wave functions, only three FFs remain, which do
not necessarily require orbital angular momenta different from
zero. In the covariant decomposition of the hadronic matrix
elements they show up as coefficients of $\gamma^+$, $\gamma^+
\gamma^5$ and $i\sigma^{+j}$, respectively. Taking only into
account these dominant form factors we end up with comparably
simple expressions for the $p\bar{p} \, \rightarrow \, \Lambda_c
\bar{\Lambda}_c$ helicity amplitudes. See Eqs.~(3.11)-(3.15) in
Ref.~\cite{Goritschnig:2008tt}. A more detailed account of these
results and their derivation will be given in a forthcoming
publication~\cite{Goritschnig:2008}.
\section{Modelling the GPDs}
In the forward hemisphere, well above production threshold, we
have to consider only the DGLAP region where, according to
Ref.~\cite{Diehl:2000xz}, the GPDs can be constructed as overlap
of light cone wave functions. This is due to the kinematical
requirement for the production of a heavy $c\bar{c}$ pair in the
hard subprocess, in which the virtuality of the gluon has to be at
least $4m_c^2$. For the $\Lambda_c$ it is certainly a good
approximation to consider only its valence Fock state. For the
proton higher Fock states may be important, but the required
overlap with the $\Lambda_c$ projects out only appropriate
spin-flavor combinations of its valence Fock state. In addition,
heavy quark effective theory (HQET) implies that, up to
corrections in the inverse of the heavy quark mass, the light
degrees of freedom in the $\Lambda_c$ couple to spin and isospin
zero so that the $c$ quark carries the helicity of the
$\Lambda_c$~\cite{Isgur:1990pm}. In this case the three form
factors which remain, if non-zero angular momentum contributions
in the wave functions are neglected, become even identical.

Since the light quarks in the $\Lambda_c$ act only as spectators
it is tempting to replace them by a single particle, namely a
(scalar, isoscalar) diquark, and to model both, the $\Lambda_c$
and the proton, as quark-diquark bound states. Only the quark
becomes active and undergoes a hard scattering. The quantum
numbers of the spectators are subsumed in the diquark. In
Ref.~\cite{Kroll:1988cd} $p \bar{p} \, \rightarrow \, \Lambda_c
\bar{\Lambda}_c$ has already been studied within a quark-diquark
model, but without using the general framework of GPDs. For our
present investigation we take a quark-diquark wave function for
the $\Lambda_c$  which is similar to the one used for a
calculation of heavy-baryon transition form
factors~\cite{Korner:1992uw}:
\be \KET{\Lambda_c^+ : \pm} \sim \KET{c_{\pm} S_{\left[ud\right]}}
\quad \text{with} \quad \Psi_{\Lambda}(x,k_\perp)  = N_{\Lambda}
\, \left(1-x\right) \,
e^{-\frac{a_{\Lambda}^2}{x\left(1-x\right)}\VEC{k}_{\perp}^2} \,
e^{ - a_{\Lambda}^2 M^2
\frac{\left(x-x_0\right)^2}{x\left(1-x\right)}}.
\label{Eq_LambdaWaveFct} \ee
The momentum fraction carried by the heavy quark is denoted by
$x$, its transverse momentum component by $\VEC{k}_{\perp}$. The
relevant quark-diquark wave function of the proton, on the other
hand, is chosen similar to the one in Ref.~\cite{Kroll:1988cd}:
\be \KET{p: \pm} \sim \KET{u_{\pm} S_{\left[ud\right]}}
\quad \text{with} \quad \Psi_p(x,k_\perp)  = N_p \,
\left(1-x\right) \, e^{-
\frac{a_p^2}{x\left(1-x\right)}\VEC{k}_{\perp}^2}.
\label{Eq_ProtonWaveFct} \ee
%
%\footnote{$x$ denotes the momentum fraction carried by the quark.}
%
These are pure $s$-wave wave functions with the helicity of the
active quark being identical with the helicity of the baryon.
Thus, according to the foregoing remarks, only three form factors
are non-zero and they become even identical. For the model wave
functions (\ref{Eq_LambdaWaveFct}) and (\ref{Eq_ProtonWaveFct})
the overlap integral can then be done analytically and the form
factor can immediately be evaluated from the GPD.
\section{Numerical Results and Conclusions}
For our numerical calculations the transverse size parameters
$a_\Lambda=a_p=1.1$~GeV$^{-1}$ are chosen in such a way that they
provide a value of $\approx 280$~MeV for the mean intrinsic
transverse momentum $<k_\perp^2>^{1/2}$ of a quark inside the
proton. The wave-function normalizations are fixed such that the
probabilities to find the quark-diquark states in a proton and a
$\Lambda_c$ are $0.5$ and $0.9$, respectively.

\begin{figure}[hbt]
   \epsfig{file=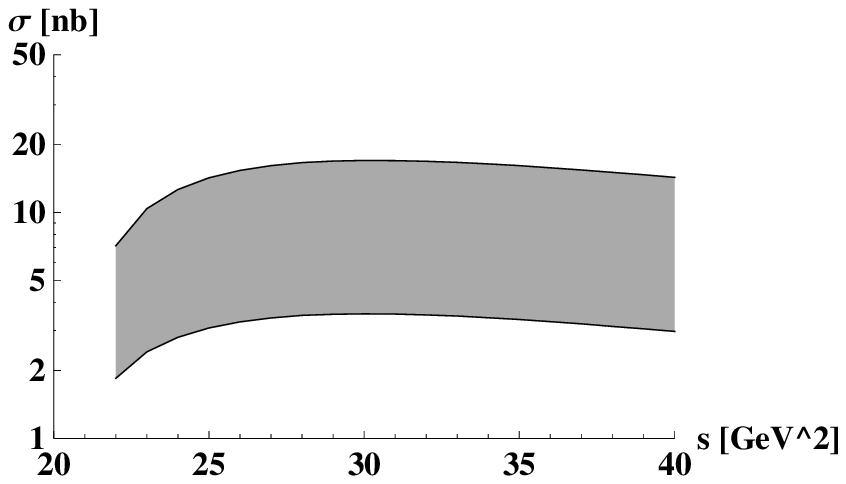,height=4.3cm,clip=}
   \hfill
   \epsfig{file=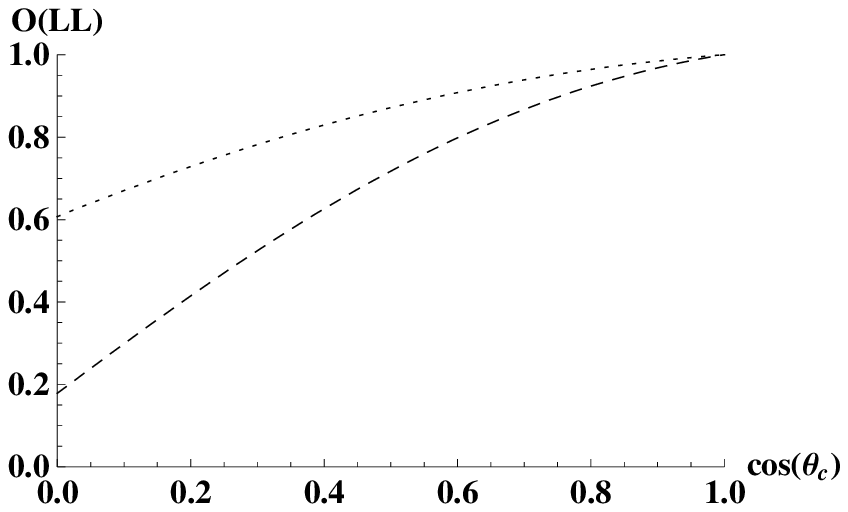,height=4.3cm,clip=}
   \vspace{-0.3cm}
   \caption{{\em Left:} The $p  \bar{p} \, \rightarrow
   \, \Lambda_c \bar{\Lambda}_c$ integrated cross section $\sigma$
   versus Mandelstam $s$ for a simple quark-diquark model.
  {\em Right:} The corresponding spin-spin correlation functions
  $-C_{LL}$ and $D_{LL} = - K_{LL}$ (dashed and dotted lines, respectively).}
    \label{Fig_CrossSection}
\end{figure}

The integrated cross section for  $p \bar{p} \, \rightarrow \,
\Lambda_c \bar{\Lambda}_c$ is plotted in
Fig.~\ref{Fig_CrossSection}. It is of the order of $\mathrm{nb}$,
which is still in the range of high precision experiments and
comparable in magnitude with the result in
Ref.~\cite{Kroll:1988cd}. Shown is a band which indicates the
uncertainty in the integrated cross section caused by varying the
transverse size parameters $a_\Lambda$ and $a_p$ from $0.75$~GeV$^{-1}$ to
$1.1$~GeV$^{-1}$. If the probabilities to find the quark-diquark states in a
proton and a $\Lambda_c$ are scaled by factors $\pi_p$ and
$\pi_\Lambda$, respectively, the cross section has to be
multiplied by $(\pi_p\, \pi_\Lambda)^2$.

Single spin observables are zero since all our amplitudes are
real. But we have various non-vanishing spin-spin correlation
functions.
Generically a spin-spin correlation function is defined as
\be \mathcal{O}_{ij} = \frac{\sigma\left(ij\right)+
\sigma\left(-i-j\right)-\sigma\left(i-j\right)-\sigma\left(-ij\right)}
{\sigma\left(ij\right)+\sigma\left(-i-j\right)+
\sigma\left(i-j\right)+\sigma\left(-ij\right)}, \ee
with $\sigma\left(ij\right)$, $i,j = L,N,S$ denoting cross
sections for the case that two of the particles participating in
the scattering process are polarized into longitudinal, normal (to
the scattering plane) or sideways direction, respectively
($L\,\bot\, N\bot\, S$). Such spin-spin correlation functions are,
e.g., $A_{ij}$ (initial spin correlations), $C_{ij}$ (final spin
correlations), $D_{ij}$ (polarization transfer from $p$ to
$\Lambda_c$), $K_{ij}$ (polarization correlation between $p$ and
$\bar{\Lambda}_c$).

Because we have only one GPD in our model the spin-spin
correlation functions of the process $p\bar{p} \rightarrow
\Lambda_c\bar{\Lambda}_c$ are equal to the ones of the subprocess
$u\bar{u} \rightarrow  c\bar{c}$, since the form factor appears as
a prefactor in the cross sections and cancels when calculating the
spin correlators. The longitudinal spin-correlation functions
$-C_{LL}$ and $D_{LL} = - K_{LL}$ are shown in
Fig.~\ref{Fig_CrossSection}. Since helicity is conserved if the
gluon couples to a light quark, proton and antiproton are forced
to have opposite helicities and hence $A_{LL}=-1$.

Non-trivial spin-effects for $p\bar{p} \rightarrow
\Lambda_c\bar{\Lambda}_c$, which differ from those of the
$u\bar{u} \rightarrow  c\bar{c}$ subprocess, can, e.g., be
achieved by taking more sophisticated wave functions for the
proton and the $\Lambda_c$. One could, e.g., think of an admixture
to the $\Lambda_c$ wave function in which the helicity of the c
quark is opposite to the $\Lambda_c$ helicity. This would lead to
three different form factors, which do not cancel anymore in the
spin-spin correlation functions.

Since the constraints posed by HQET (in the limit $m_c \rightarrow
\infty $) are already implemented in the simple quark diquark
model, we do not expect that the use of three-quark wave functions
for the proton and the $\Lambda_c$ will change the results
dramatically. Details of the wave functions will most likely show
up in spin observables. Existing models for the three-quark
light-cone wave function of the proton~\cite{Bolz:1996sw},
~\cite{Bartels:2007aa} get some support from their success in
phenomenological applications and from lattice
QCD~\cite{Gockeler:2008xv}, but much less is known about the
$\Lambda_c$ wave function. The sensitivity of unpolarized and
polarized cross sections on the model for the $\Lambda_c$ wave
function is presently under investigation
~\cite{Goritschnig:2008}. It would be interesting to see
experimentally, whether polarization phenomena in $p\bar{p}
\rightarrow \Lambda_c\bar{\Lambda}_c$ are governed by the c-quark
polarization, as in our simple quark-diquark model, or whether
other wave function components will be necessary.
%
%
%
%%%%%%%%%%%%%%%%%%%%%%%%%%%%%%%%%%%%%%%%%%%%%%%%
%% BACKMATTER
%%%%%%%%%%%%%%%%%%%%%%%%%%%%%%%%%%%%%%%%%%%%%%%%

\begin{theacknowledgments}
  A.T.G. acknowledges the support of the \lq\lq Fonds zur F\"orderung der
  wissenschaftlichen Forschung in \"Osterreich\rq\rq\, (project DK W1203-N08).
\end{theacknowledgments}



\bibliographystyle{aipproc}   % if natbib is available

%\bibliography{sample}

%  \typeout{******************************************}
%  \typeout{** Please run "bibtex \jobname" to optain}
%  \typeout{** the bibliography and then re-run LaTeX}
%  \typeout{** twice to fix the references!}
%  \typeout{******************************************}

%\end{document}

%%%%%%%%%%%%%%%%%%%%%%%%%%%%%%%%%%%%%%%%%%%
%% The following lines show an example how to produce a bibliography
%% without the help of the BibTeX program. This could be used instead
%% of the above.
%%%%%%%%%%%%%%%%%%%%%%%%%%%%%%%%%%%%%%%%%%%

\end{document}